\documentclass[conference, a4paper]{IEEEtran}
\IEEEoverridecommandlockouts

\usepackage{cite}
\usepackage{todonotes}
\usepackage{url}

\ifCLASSINFOpdf
  \usepackage{graphicx}

\else

\fi

\usepackage[cmex10]{amsmath}
\usepackage{multirow}
\usepackage{array}
\usepackage[lofdepth,lotdepth]{subfig}

\AtBeginDocument{%
  \renewcommand\abstractname{Abstract}
}

\begin{document}

\IEEEpubid{\makebox[\columnwidth]{978-1-6654-3649-6/21/\$31.00 ©2021 IEEE\hfill}
\hspace{\columnsep}\makebox[\columnwidth]{}}

%
% paper title
% can use linebreaks \\ within to get better formatting as desired
\title{Semi-Supervised Segmentation of  Multi-vendor and Multi-center Cardiac MRI using Histogram Matching}

% author names and affiliations
% use a multiple column layout for up to three different
% affiliations

\author{\IEEEauthorblockN{Mahyar Bolhassani}
\IEEEauthorblockA{Biomedical Engineering Department\\
Istanbul Technical University, Turkey\\
bolhassani19@itu.edu.tr}
\and
\IEEEauthorblockN{Ilkay Oksuz}
\IEEEauthorblockA{ Computer Engineering Department\\
Istanbul Technical University, Turkey\\
oksuzilkay@itu.edu.tr
}
}

% make the title area
\maketitle

\begin{abstract}

Automatic segmentation of the heart cavity is an essential task for the diagnosis of cardiac diseases. In this paper, we propose a semi-supervised segmentation setup for leveraging unlabeled data to segment Left-ventricle, Right-ventricle, and Myocardium. We utilize an enhanced version of residual U-Net architecture on a large-scale cardiac MRI dataset. Handling the class imbalanced data issue using dice loss, the enhanced supervised model is able to achieve better dice scores in comparison with a vanilla U-Net model. We applied several augmentation techniques including histogram matching to increase the performance of our model in other domains. Also, we introduce a simple but efficient semi-supervised segmentation method to improve segmentation results without the need for large labeled data. Finally, we applied our method on two benchmark datasets, STACOM2018, and M\&Ms 2020 challenges, to show the potency of the proposed model. The effectiveness of our proposed model is demonstrated by the quantitative results. The model achieves average dice scores of 0.921, 0.926, and 0.891 for Left-ventricle, Right-ventricle, and Myocardium respectively.

%\boldmath
\end{abstract}

\IEEEpeerreviewmaketitle

\IEEEpubidadjcol

\section{Introduction}
According to World Health Organization (WHO), the most prominent cause of mortality among people worldwide is cardiovascular diseases \cite{Chen}. Accurate and early detection of cardiac disease can be instrumental in the treatment plan. Modern medical imaging techniques provide huge improvements in the diagnosis of cardiovascular-related diseases \cite{Chen}. These techniques enable us to assess the anatomical structure of the heart, both in quantitative approaches such as measuring the tumorous tissue volume, and qualitative ways like deblurring images. There are some challenges in medical images that need to be addressed appropriately such as having a small portion of the image as the area of interest and domain shift.

One of the most challenging problems in the segmentation of medical images is the lack of annotated or labeled data. The annotation process is both a costly and tedious task. This process demands hiring medical experts for annotation which makes engineers search for methods and architectures to reduce the need for large labeled data. Petitjean et al. \cite{Petitjean} reviews the cardiac image segmentation algorithms using 2D models. The fact that cardiac images are 4D, 3 spatial dimensions accompanied by temporal dimension, it gives researchers large numbers of 2D labeled images of the heart. 

Poudel et al. \cite{Poudel} proposed to leverage the recurrent neural network approach to segment cardiac images. Ronneberger et al.\cite{Ronneberger} introduced U-Net architecture for image segmentation which has been widely used for segmentation tasks. He et al. \cite{He} proposed ResNet architecture in which the residual connections enhanced the performance of the neural network in computer vision classification and segmentation tasks. This architecture is widely used in medical image segmentation and authors in \cite{Kerfoot} applied this architecture to achieve their goal. Zhang et al. introduced residual U-net architecture for road area extraction \cite{Zhang}. In this work, we have used both U-net and residual U-net for the supervised learning scheme. Bai et a. \cite{Bai} came up with a semi-supervised approach to increase the input size without the need for more labeled data. Although their approach improved the performance of the model especially when the numbers of labeled data are limited, early noisy predictions can adversely affect the performance of the model. Therefore, the search for a better semi-supervised method is still an ongoing research field.  
% \todo[inline]{IO to MB: You need to include Wenjia Bai paper to the literature review. You can include a paragraph on semi-supervised segmentation of computer vision (with references) images and why it is relavant if space allows}

In this work, we present a simple but effective semi-supervised method to leverage unlabeled data. In this approach, we predict labels for the unlabeled data using the trained fully supervised model. Then, retrained the model by utilizing all labeled and pseudo-labeled images.  Also, to further enhance the generalizability of the proposed model and to reduce the noisy pseudo-labeling predictions, we apply an augmentation method named histogram matching which helps the training data to mimic the intensity distribution of the unseen datasets. Therefore, this approach gives us the opportunity to overcome the lack of annotated data issues while improving the predictions on unseen datasets. 
Firstly, we train fully supervised models using both U-Net and residual U-Net. Then, we apply our proposed semi-supervised method to leverage the unlabeled data. Finally, we compare the performance of specific variations with the help of dice scores as the evaluation metrics.
%%% METHODS%%%%
\section{Materials and Methods}
% Automatic segmentation of cardiac MRI images is a challenging task which needs a considerable number of annotated data. The problem is that the scans usually are done in different centers (hospitals, countries) or with various vendors (Siemens, Canon, etc). Therefore, it generates various alignments or completely different intensity distributions even for a single patient. Thus, our automatic model should address those variations in order to successfully perform a robust segmentation task. As mentioned before, we don't have the access to all kinds of scans with their annotated masks.

In this section, we introduce the benchmark datasets, data augmentation strategies, and model architectures. To overcome the issues of domain shift, and lack of enough annotated samples, we performed data augmentation techniques to enhance the generalization of our model.  Finally, we elaborate on the semi-supervised method.

\subsection{Benchmark datasets}
In this study, two different datasets are utilized to assess the performance of the segmentation model. The first benchmark dataset consists of cardiac image scans which are Multi-Centre (different countries), Multi-Vendor (various vendors), and Multi-Disease of the challenge (M\&Ms) \cite{MnMS}.
As training data, 150 labeled images are prepared from two different vendors (75 each) Also, there are 25 unlabeled images from a third vendor which in this research, we considered as the unlabeled data for the semi-supervised method. Clinician experts have segmented cardiac images including three important structures: annotations for the left ventricle (LV), right ventricle (RV), and myocardium (MYO) regions which each respectively indicates labels of 1 for LV, 2 for MYO, and 3 for RV. 
The second benchmark dataset is from Full Left Ventricle Quantification Challenge(LVQuan). The dataset includes 145 cardiac MRI subjects from a clinical environment. For each subject, 20 frames included the whole cardiac cycle from systole to diastole. For all the above images, they have provided all ground truth values of the LV indices for every single frame \cite{STACOM2018}.

\subsection{Data augmentation and histogram matching}
\begin{figure}[h]
	\centering
	% \shorthandoff{}  %\usepackage[turkish]{babel} kullanıldığı için komuta gerek var
	\includegraphics[width=\linewidth]{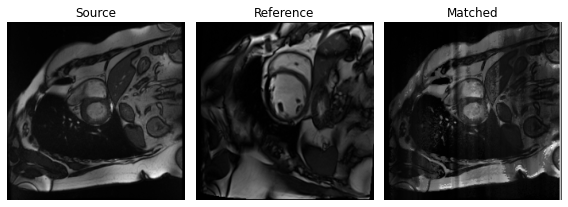}
	% \shorthandon{} %\usepackage[turkish]{babel} kullanıldığı için komuta gerek var
	\caption{Histogram matching result. From the left to right respectively: the source image, reference image, the matched image}
	\label{fig:hist_M}
\end{figure}

A large number of training samples is crucial to enhance the performance of a deep model. Using data augmentation can alleviate the issue of lack of a large number of labeled data by bringing variations to the training dataset which can significantly improve the performance of the model on unseen samples. In this work, we have applied several augmentation techniques including rotation of -45 to 45-degree, -90 to 90-degree rotation, horizontally flipping with the rate 50\% of the images,  and sharpening. 
Another simple but highly effective augmentation technique that we have applied, is histogram matching \cite{Ma}. In the first dataset, there is a set of data from the third vendor which is unlabeled. To enhance our supervised model, we augmented the labeled datasets from vendor A and vendor B using the intensity distribution of the third vendor. In this method, the histogram of vendors A and B is matched to the third vendor dataset. Thus, the augmented training data can gain some knowledge about unlabeled data. As a result, the model can make better predictions on them and avoid mislabeling samples from vendor C. Figure \ref{fig:hist_M} illustrates the histogram matching on a sample image from vendor B of the training dataset as a source image via referencing an image randomly sampled from vendor C.
\subsection{Architecture}
We have utilized U-Net and Residual U-Net for the task of cardiac MR segmentation. Having these two architectures as benchmarks will allow us to evaluate the performance of the semi-supervised segmentation algorithm on the datasets.
\subsubsection{U-Net}
U-Net is a state-of-the-art convolutional neural network solution for the segmentation task. Based on \cite{Ronneberger} U-Net has a U-shape and is symmetric. It consists of two major parts: the encoder consists of 3 × 3 convolution layers each follows by a ReLU activation function(contracting path), and the decoder (expansive path). For downsampling input images in the encoder, a 2 × 2 max-pooling operation (stride 2) is applied. In contrast, in the upsampling path, a 2 × 2 convolution layer is used and unlike the downsampling, the number of filters in each step will halve. The connection lines between downsampling blocks and upsampling blocks enhance the local information during the segmentation task. Thus, the feature map from downsampling is concatenated with the upsampling. 
\subsubsection{Residual U-Net}
\begin{figure}[h]
	\centering
	% \shorthandoff{}  %\usepackage[turkish]{babel} kullanıldığı için komuta gerek var
	\includegraphics[width=8cm]{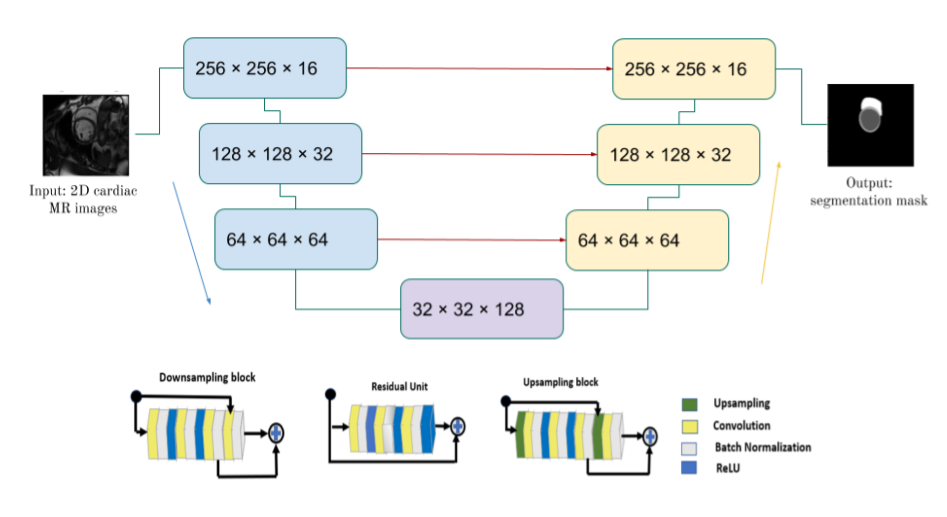}
	% \shorthandon{} %\usepackage[turkish]{babel} kullanıldığı için komuta gerek var
	\caption{Residual U-Net architecture}
	\label{fig:unet_model}
\end{figure}

The deeper a neural network becomes the more prone it is to suffer from a vanishing gradient. When we have a substantially deeper neural network, it helps to achieve more accurate results. Though the efficiency of the training will increase when they contain shorter connections between convolutional layers since it allows the layers close to the output to remember the input information. As a result, including residual blocks in each layer in the U-Net architecture improves the performance of the model and decreases the occurrence of vanishing gradients. Figure \ref{fig:unet_model}, shows the Res-U-Net architecture in detail.
\section{Semi-Supervised method}

\begin{figure}[h]
	\centering
	% \shorthandoff{}  %\usepackage[turkish]{babel} kullanıldığı için komuta gerek var
	\includegraphics[width=8cm]{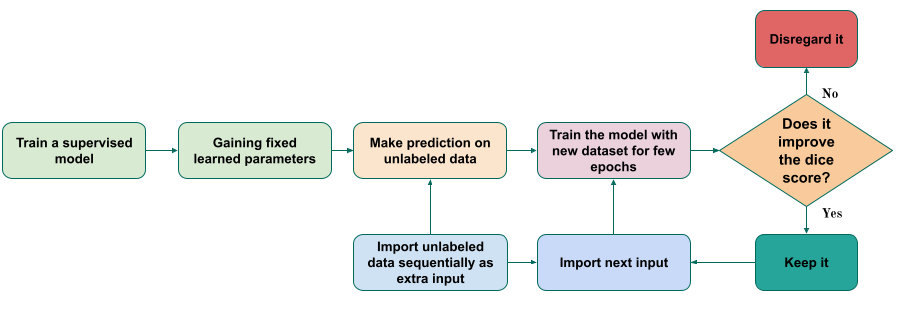}
	% \shorthandon{} %\usepackage[turkish]{babel} kullanıldığı için komuta gerek var
	\caption{The proposed semi-supervised method}
	\label{fig:ssl}
\end{figure}

% \todo[inline]{Io to MB: How many labeled cases for trianing}
In the M\&Ms dataset, we have 25 unlabeled samples which we can leverage in order to improve the generalizability of the trained model. We began with a fully supervised model (an enhanced residual U-net with data augmentation and dice loss). The predicted labels for the unlabeled data from vendor C are achieved using the trained supervised model. Due to the fact that only ES and ED time slides are annotated on the dataset, we consider ED and ES time slides for vendor C to be labeled. To avoid using low-quality annotations as we explained in section II.B, the histogram matching method has been utilized during fully supervised training. Also, we evaluate the prediction throughout a loop to check if those predictions are noisy and disregard them by the trained supervised model. Then, two datasets, the manually labeled data and the predicted labeled data are merged. The detailed implementation of the method is shown in Figure \ref{fig:ssl}. Using this method, allows the model to train on images from different vendors or centers. 

%%% EXPERIMENTAL RESULTS %%%%
\section{Implementation Details}

We trained both the U-Net and Residual U-Net architecture using the Keras library on NVIDIA Quadro GV100 GPU. The training took 3 hours and 48 minutes (28 seconds for each epoch) for 300 epochs for the Residual U-Net architecture, though the U-Net architecture was significantly slower, taking 5 hours (33 seconds for each epoch.) We implemented our model in a 2D fashion so that we needed to load the input data slice by slice in End systole (ES) and End diastole (ED), the provided annotated scans. We divided 2439 2D labeled images into 1711 training, 428 validation (20 percent of the training data), and 300 test images. To avoid having the same sequence that belongs to a certain patient in both training and testing sets during splitting data, we consider the number of 2D images of each patient (z-axis, and time-axis slides are different) to assure correct splitting. This splitting process is essential to avoid over-fitting during training. In addition, we used 478 2D images of the unlabeled data. Because of inconsistency during data acquisition, we need to apply a preprocessing approach to ensure the model receives uniform input data. In this work, we cropped the images with 224 by 224 pixels using the center point of each slice. In the next step, normalization for the pixel values of each image has been done in both the training set and test set separately to avoid picking phenomena. We choose an ADAM optimizer with an initialization of the learning rate equal to 0.001. 
In addition, we used both categorical cross entropy and dice loss for the training of both U-Net and Residual U-Net architectures throughout the back-propagation.  We also used the average dice coefficient as our evaluation metric.
In Equation \ref{eq:eq2} A represents the ground truth image (segmentation mask) and B refers to the predicted segmentation mask.
% as described in Equation \ref{eq:eq1}.
% \begin{align}
% CE = -\sum_{i=1}^{C'=2}t_{i} log (s_{i}) = -t_{1} log(s_{1}) - (1 - t_{1}) log(1 - s_{1})
% \label{eq:eq1}
% \end{align} 
% \todo[inline]{MB. what is t, what is s..Please explain each variable} 
\begin{align}
DC = \frac{2\left | A\bigcap B\right|}{\left | A \right | + \left | B \right |}
\label{eq:eq2}
\end{align}

\section{Experimental Results}
In this section, we provide quantitative and qualitative results of the proposed supervised and semi-supervised segmentation frameworks.
\subsection{Quantitative Results}
The average dice scores of all three regions of interest, (LV, RV, and Myocardium) on training, validation, and test sets regards to fully supervised scenarios are shown in Table \ref{tab:1} (STACOM LVQuan dataset), and Table \ref{tab:2} (M\&Ms dataset). In this step, we used categorical cross-entropy loss. In Table \ref{tab:1}, we do not report results on the RV region since the dataset just has annotations on LV and Myocardium.
%  \todo[inline]{IO to MB: can we include standard deviation numbers??? MAKE TGE BEST RESULT BOLD}
\begin{table}[htbp]
  \centering
  \caption{Results for STACOM LVQuan 2018 dataset}
    \begin{tabular}{ccccc}
    \hline
    % \multirow {} & \multicolumn{2}{c}{U-Net} &\multicolumn{2}{c}{Res-U-Net}  \\
    %     \cline{2-5}

     DSC & no-aug. & with aug. & no-aug &  with aug. \\
    \hline
    Tr-LV     & 0.784  & 0.778 & 0.985    & 0.983    \\
    Val-LV     & 0.768  & 0.772 & 0.983   & 0.985    \\
    Tst-LV     & 0.748  & 0.761 & 0.941   & \textbf{0.957}   \\
    Tr-Myo     & 0.842 & 0.868 &  0.992   & 0.994   \\
    Val-Myo     & 0.802 & 0.832 & 0.962   & 0.962   \\
    Tst-Myo     & 0.803 & 0.815 & 0.923   &\textbf{0.942}   \\
    \hline

    \end{tabular}%
  \label{tab:1}%
\end{table}%

\begin{table}[htbp]
  \centering
  \caption{Results for M\&Ms challenge}
    \begin{tabular}{ccccc}
    \hline
    % \multirow {} & \multicolumn{2}{c}{U-Net} &\multicolumn{2}{c}{Res-U-Net}  \\
    %     \cline{2-5}

     DSC & no-aug. & with aug. & no-aug &  with aug. \\
    \hline
    Tr-LV     & 0.781  & 0.769 & 0.921    & 0.914   \\
    Val-LV     & 0.725  & 0.751 & 0.82   & 0.878    \\
    Tst-LV     & 0.702  & 0.711 & 0.795   & \textbf{0.851}   \\
    Tr-RV     & 0.74  & 0.742 & 0.866   & 0.872    \\
    Val-RV     & 0.656 & 0.674 & 0.705  & 0.791   \\
    Tst-RV     & 0.623 & 0.661 & 0.683   & 0.776  \\
    Tr-Myo     & 0.787 & 0.766 & 0.747  & 0.761   \\
    Val-Myo     & 0.697 & 0.721 & 0.703  & 0.73   \\
    Tst-Myo     & 0.614 & 0.622 & 0.638   & \textbf{0.648}   \\
   \hline

    \end{tabular}%
  \label{tab:2}%
\end{table}%

Based upon the results shown in Tables \ref{tab:1} and \ref{tab:2}, we can see that Residual U-Net with augmentation technique outperforms the vanilla U-net and Res-net without augmentation. Therefore, Residual U-Net with augmentation is chosen as our architecture for the supervised model in order to predict masks of the unlabeled dataset. We trained the model based on the defined hyperparameters for 500 epochs and dice loss while applying regular augmentation techniques and the results are shown in Table \ref{tab:3}. We have considered 8 scenarios containing, FS (fully supervised), FS50 (FS with \%50 of labeled data), FSH (FS with histogram matching), FS50H (FS with \%50 of labeled data and histogram matching), and the same scenarios for SS (semi-supervised).
The results indicate that the proposed semi-supervised method improves the dice scores compared to segmentation without the unlabelled cases. It is observed that the dice score of semi-supervised with histogram matching for the validation set is at the highest.  
% \todo[inline]{IO to MB. Write one sentence on why do you think  the performance increased..}
\begin{table}[htbp]
  \centering
  \caption{Results of the semi-supervised method for labeled and unlabeled M\&Ms challenge data}
    \begin{tabular}{ccccccc}
    \hline
    % \multirow {Dice} & \multicolumn{2}{c}{U-Net} &\multicolumn{2}{c}{Res-U-Net}  \\
        % \cline{2-5}

    DSC & Tr-LV & Val-LV & Tr-RV &  Val-RV & Tr-Myo & Val-Myo \\
    \hline
    FS & 0.938  & 0.779 & 0.93 & 0.810 & 0.893 & 0.706  \\
    FS50 & 0.913  & 0.77 & 0.901 & 0.787 & 0.879 & 0.690 \\
    FSH & 0.963  & 0.782 & 0.951 & 0.793 & 0.904 & 0.714  \\
    FS50H & 0.941  & 0.773 & 0.936 & 0.733 & 0.911 & 0.714  \\
    SS & 0.966  & 0.873 & 0.96 & 0.892 & 0.899 & 0.778  \\
    SS50 & 0.950  & 0.801 & 0.941 & 0.821 & 0.912 & 0.765  \\
    SSH & \textbf{0.971}  & \textbf{0.921} & \textbf{0.977} & \textbf{0.926} & \textbf{0.946} & \textbf{0.891} \\
    SS50H & 0.951 & 0.851 & 0.968 & 0.858 & 0.944 & 0.839  \\
  \hline
    \end{tabular}%
  \label{tab:3}%
\end{table}%

\subsection{Qualitative Results}
\begin{figure}[htb]
	\centering
	\includegraphics[width=9cm]{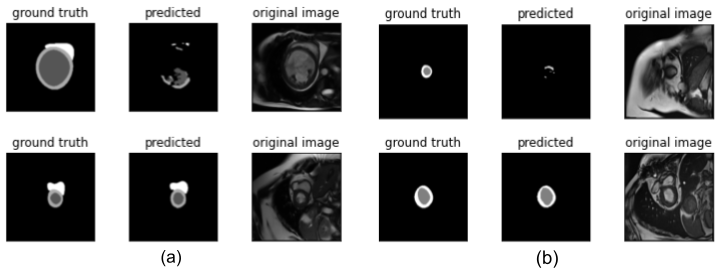}
	\caption{Semi-supervised results considering labeled and unlabeled M\&Ms challenge data. first row: worst results, and second row: best results. (a) SF results, (b) SS results. }
	\label{fig:BWR}
\end{figure}

\begin{figure}[htb]
	\centering
	\includegraphics[width=9cm]{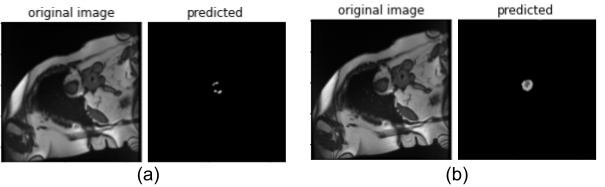}
	\caption{predicted mask for an image in vendor C (a)  using FS model, (b) using FSH model.}
	\label{fig:Pun}
\end{figure}

The results of our semi-supervised approach, are illustrated in Figure \ref{fig:BWR}. In this figure, the first row indicates the worst results and the second row shows the best ones;(a) shows SF results while (b) illustrates SS results. The impact of applying histogram matching in the prediction of labels for vendor C samples is shown in Figure \ref{fig:Pun}. It is noticeable that after applying histogram matching, the model can predict more confident predictions on the unlabeled samples which is an indicator of a robust model. 
% \todo[inline]{IO: please comment on the results.}

\section{Conclusion and future studies}

Our work focused on the segmentation of the myocardial cavity on two benchmark datasets with two states of the network architectures. We proposed supervised and semi-supervised segmentation methods to increase the segmentation performance utilizing the information in the unlabeled data. The residual U-Net architecture with data augmentation setup outperforms other scenarios. Thus, we focused on residual U-net setup for the semi-supervised method with the help of the histogram matching technique to showcase the increased performance on M\&Ms dataset.

We introduced a simple but effective semi-supervised method to utilize the unlabeled data to enhance the predictions on unseen data. Histogram matching showed significant improvements in the performance of the semi-supervised model since the training dataset has a chance to mimic the intensity distribution of the unlabeled data. Thereafter, we predicted the segmentation of the unlabeled dataset and then appended them to the main training set. Even though with semi-supervision, we reduced the impact of adverse predictions on our model to some extent, one limitation of our method could be the inaccurate segmentation achieved on the unlabeled dataset.  In future work, we aim to validate our technique in a larger cohort with more variations (unlabeled data is from a different vendor). We aim to train our framework in a continuous learning scheme in datasets, where manual annotations are limited. In addition, we consider applying label-propagation for unlabeled slides of each patient to improve the semi-supervised method.

%it method has a limitation which is if the first prediction is not accurate the error function will increase as the training continuing. 

%%% CONCLUSIONS and FUTURE WORK %%%%
\section*{Acknowledgments}
This paper has been produced benefiting from the 2232 International Fellowship for Outstanding Researchers Program of TUBITAK (Project No: 118C353). However, the entire responsibility of the paper belongs to the owner of the paper. The financial support received from TUBITAK does not mean that the content of the publication is approved in a scientific sense by TUBITAK.
%This paper has been produced benefiting from the 2232 International Fellowship for Outstanding Researchers Program of TUBITAK (Project No: 118C353). However, the entire responsibility of the publication/paper belongs to the owner of the paper. The financial support received from TUBITAK does not mean that the content of the publication is approved in a scientific sense by TUBITAK.}

\section{Appendix}
\begin{figure}[htb]
	\centering
	\includegraphics[width=9cm]{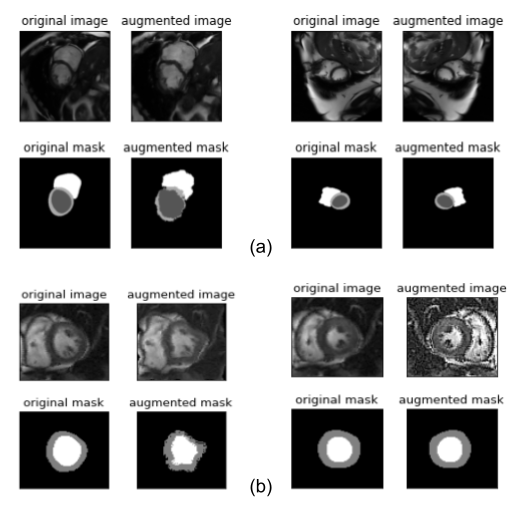}
	\caption{Some sample augmentation images. }
	\label{fig:aug}
\end{figure}

\begin{figure}[htb]
	\centering
	\includegraphics[width=8cm]{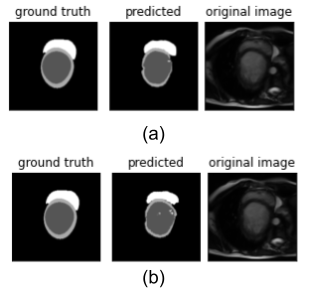}
	\caption{Some sample images after applying attention module. }
	\label{fig:att}
\end{figure}

\begin{figure}[htb]
	\centering
	\includegraphics[width=9cm]{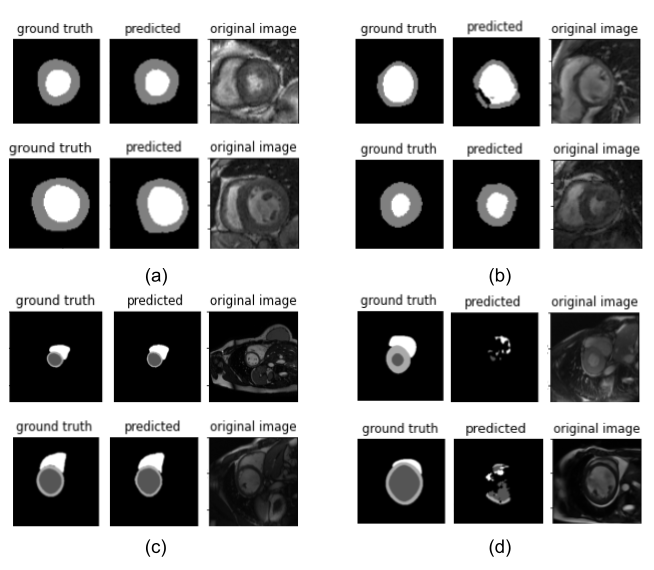}
	\caption{Some samples of the worse and best results of different fully supervised models that we explored. }
	\label{fig:FS}
\end{figure}

Figure \ref{fig:aug} shows a few augmented images that we have done in our implementation. 
We also applied an attention module to improve the performance of our model. Figure \ref{fig:att} illustrates a few sample results. 
Also, Figure \ref{fig:FS} explores more sample images that illustrate best and worse predictions using different fully supervised models.
\end{document}